\documentclass[twoside,12pt,reqno]{amsart}
\usepackage{a4wide}
\usepackage{amsmath}
\usepackage{amsfonts}
\usepackage{amssymb}
\usepackage{pstricks,pst-node,pst-text,pst-3d}
\usepackage{times}
\usepackage{graphics}

\theoremstyle{plain} 

\psset{framearc=0.0,framesep=0.0truecm,nodesep=3pt,unit=0.75truecm,%
arrowsize=4pt 2,arrowinset=0.4}

\newcommand{\ed}{\end{document}}
\newcounter{mycnt}[section]

\def\one{\hbox{{1}\kern-.25em\hbox{l}}}

\def\!{\kern -0.15ex}

\begin{document}
\title[{The mixed two qutrit system}]{The mixed two qutrit system: local unitary invariants, entanglement monotones, and the SLOCC group $SL(3,{\mathbb C})$.}
\author[P.~D.~ Jarvis]{P.~D.~ Jarvis$^\dagger$ \\
School of Mathematics and Physics, University of Tasmania
}
\address{School of Mathematics and Physics, University of Tasmania, 
Private Bag 37 GPO, Hobart Tas 7001, Australia. 
\hfill \mbox{~~}$^{\dagger}$Alexander von Humboldt Fellow}


\begin{abstract}
\mbox{}\\
We consider local unitary invariants and entanglement monotones for the mixed two qutrit system. Character methods for the local $SU(3)\times SU(3)$ transformation group are used to establish the count of algebraically independent polynomial invariants up to degree 5 in the components of the density operator. These are identified up to quartic degree in the standard basis of Gell-Mann matrices, with the help of the calculus of $f$ and $d$ coefficients. Next, investigating local measurement operations, we study a SLOCC qutrit group, which plays the role of a `relativistic' transformation group analogous to that of the Lorentz group $SL(2,{\mathbb C})_{\mathbb R}\simeq SO(3,1)$ for the qubit case. This is the group $SL(3,{\mathbb C})_{\mathbb R}$, presented as a group of real $9\times 9$ matrices acting linearly on the 9-dimensional space of projective coordinates for the qutrit density matrix. The counterpart, for qutrits, of the invariant $4\times 4$ Minkowski metric of the qubit case, proves to be a certain $9\times 9 \times 9$ totally symmetric three-fold tensor generalizing the Gell-Mann $d$ coefficient. Using this structure, we provide a count of the corresponding LSL polynomial invariants using group character methods. Finally, we give an explicit construction of the lowest degree quantity (the cubic invariant) and its expansion in terms of $SU(3)\times SU(3)$ invariants, and we indicate how to construct higher degree analogues. These quantities are proven to yield entanglement monotones. This work generalizes and partly extends the paper of King, Welsh and Jarvis (\emph{Journal of Physics A}, 40(33):10083, 2007) on the mixed two qubit system, which is reviewed in an appendix.\\[.2cm]
\vfill
\end{abstract}
\vfill
\normalsize
\date{Dec 2013}
\maketitle
%
\pagebreak
\section{Introduction}
\label{sec:Intro}
The quantification of entanglement in composite quantum systems remains a central question of foundational importance, as well as in terms of practical understanding and implementation of protocols in quantum information and quantum computation. We refer to \cite{HorodeckiEtAl:2009qe} for a comprehensive review of entanglement, and \cite{BengtssonZyczkowski2006bkgqs} for a modern text. Beyond general entropic measures, which typically require optimization over arbitrary 
realizations or preparations of a quantum state, much analysis has been devoted to concretely defined invariant quantities, which are polynomials in the coordinates parametrizing the state. Many results in this direction concern multipartite qubit systems, ranging from detailed studies for low numbers of subsystems (up to five or six), to  systematic identifications of classes of invariants which are available for the general case of $K$ subsystems (specific studies will be cited in the main text). The former category typically treats pure states, and there are fewer such results available for mixed systems, where a density matrix rather than a state vector must be adopted for the description of the quantum state.

Notwithstanding the central role of two state models in quantum mechanics and the focus given to qubits in quantum information and quantum computation, there is also considerable interest in the next simplest qu$\!\!D\!$it system, namely the qutrit ($D=3$). 
Three state systems (involving, for example, a working qubit and an ancilla state) may in fact underlie experimental  implementations of quantum protocols \cite{VaziriWeihsZeilinger2002tde,YuYiSongMei2008p2qut}, and 
it is of great physical importance to characterize their states and interactions. Alongside the enormous effort on qubit systems, there is a smaller literature on qutrits, such as recent studies of the geometry of qutrit states (the generalization of the Bloch sphere for example \cite{goyalsimonsingh2011gbsqut,SarbickiBengtsson2013dqt}). More fundamentally, composite qutrit systems obviously provide a further challenge to the elucidation of entanglement, and the understanding of entanglement measures \cite{DerkaczJakobczyk2007e2qut}.

The present paper presents results on polynomial invariants and entanglement monotones for the two qutrit mixed system. It generalizes and partially extends our paper \cite{KingWelshJarvis2007} on the two qubit mixed system. Our results in that paper confirmed earlier enumerations of local unitary (LU) invariants for the two qubit mixed system \cite{GrasslEtAl1998cli}, and subsequent investigations of their role in separating entanglement classes \cite{makhlin2002nlp}. However, by giving a full resolution of the structure of the invariant ring, the paper provided a  complete understanding of the algebraic relations between the invariants, and their associated syzygies.   

As is well known, enumeration of LU invariants is a necessary first step towards the construction of true entanglement measures, which should also satisfy monotonicity criteria under quantum operations. By contrast with our results on the two qubit mixed system, in this paper, only partial results on local unitary invariants for the two qutrit mixed system are attained, but these are subsequently used to establish examples of \emph{bona fide} entanglement monotones, which are based on quantities polynomial in the components of the density operator.

In \S \ref{sec:LocalUnitary} below, character methods for the local $SU(3)\times SU(3)$ transformation group are used to establish the count of algebraically independent polynomial invariants up to degree 5 in the components of the density operator. These include 3 quadratic, 7 cubic and 17 quartic quantities, and they are identified up to quartic degree in the standard basis of Gell-Mann matrices, with the help of the calculus of $f$ and $d$ coefficients (Tables \ref{tab:QuadCub}, \ref{tab:GradedCount}). Going in \S \ref{sec:SL3C} to local measurement operations, we study a SLOCC qutrit group, which plays the role of a `relativistic' transformation group analogous to that of the Lorentz group $SL(2,{\mathbb C})_{\mathbb R}\simeq SO(3,1)$ for the qubit case. This is the local special linear (LSL) group $SL(3,{\mathbb C})_{\mathbb R}$, presented as a group of real $9\times 9$ matrices, acting linearly on the 9-dimensional space of projective coordinates for the qutrit density matrix. The counterpart, for qutrits, of the invariant $4\times 4$ Minkowski metric of the qubit case, proves to be a certain $9\times 9 \times 9$ totally symmetric three-fold tensor generalizing the Gell-Mann $d$ coefficient. This 9-dimensional matrix group presentation of $SL(3,{\mathbb C})_{\mathbb R}$ is here denoted $H_d(8,1)$ by analogy with the Lorentz case. We prove directly the isomorphism of Lie algebras, $h_d(8,1)\cong sl(3,{\mathbb C})_{\mathbb R}$. We provide a count of the corresponding two qutrit mixed state LSL polynomial invariants using group character methods. These quantities are proven to yield entanglement monotones. \S \ref{sec:SL3C} ends with the explicit identification of the two lowest degree quantities (the cubic and sextic invariants), and the expansion of the cubic in terms of the local $SU(3)\times SU(3)$ invariants is given. The paper concludes in \S \ref{sec:Conc} with a brief discussion and overview.

To aid readability in this paper, a number of definitions and derivations are relegated to several appendices.
Appendix \ref{sec:Schurology} provides a summary of the definitions and notation required for handling the group character manipulations, on which our main results are based. In particular, Theorems 1 and 2 in appendix 
\ref{sec:Schurology} respectively give counting rules for the numbers of local unitary invariants for pure and mixed state systems at each degree, and for the numbers of LSL invariants for bipatrite mixed qubit and qutrit systems at each degree. 
 Appendix \ref{sec:Monotones} presents the standard derivation for constructions of entanglement monotones based on LSL invariants, in a way which generalizes from pure states to mixed states, and from qubits to qutrits (in particular, this applies to the quantities defined in \S \ref{sec:SL3C}). For completeness, we provide in appendix \ref{sec:TwoQubitReview} a brief summary of the results of \cite{KingWelshJarvis2007} on the qubit case. There we also exemplify how the LU $SU(2)\times SU(2)$ invariants allow the construction of associated local $SL(2,{\mathbb C})\times SL(2,{\mathbb C})$ invariants, as a template for the extension to qutrits presented in  \S \ref{sec:SL3C}. 

\section{Local unitary invariants for 2 qutrit mixed states}
\label{sec:LocalUnitary}
The context for enumerating and identifying local unitary invariants which are polynomial in the coordinates defining the quantum state, is that of the representation theory of the local groups of unitary transformations acting on each subsystem. The count is given in closed form via Molien's theorem, which gives an integral representation for the Molien series $h(z)= \sum_0^\infty h_nz^n$, the generating function for the number of linearly independent invariants $h_n$ at each polynomial degree $n$. In \cite{KingWelshJarvis2007} this was evaluated for the two qubit mixed system with local $SU(2)\times SU(2)$ group, the algebraically independent invariants constructed, and the structure of the invariant ring characterized completely.  The coefficients of the Molien series  were also verified combinatorially by computations using group character methods.

In the 2 qutrit case, a pure state is described by a 9-component complex wavefunction $\psi^{i\underline{j}}$, $i,\underline{j} = 1,2,3$, while a mixed state 
density operator $\rho^{i\underline{j}, k\underline{\ell}}$ has 81 real components (subject to the constraints of positivity and unit trace). As a matrix it admits an expansion with respect to the standard basis of Gell-Mann matrices on each subspace,
\begin{equation}
\rho = \textstyle{\frac 19} {\mathbb I}_{9} + \sum_{a=1}^8 r^a \lambda_a\otimes  {\mathbb I}_{3} + 
 \sum_{\underline{a}=1}^8 {r}^{\underline{b}}  {\mathbb I}_{3}\otimes\lambda_{\underline{b}} +  
 \sum_{a, \underline{b}=1}^8 {R}^{a\underline{b}}\lambda_a\otimes \lambda_{\underline{b}},
 \end{equation}
showing that (after fixing the trace) the 80 linearly independent real components are comprised of octet vectors ${r}^{{a}}$, ${r}^{\underline{b}}$ belonging to each local $SU(3)$ transformation group, together with an $8 \times 8$ tensor\footnote{
We adopt the convention that index suffices on the subsystem spaces use the same alphabet but distinguished by an underline; thus $i, \underline{j} = 1,2,3$ in the defining representation, $a, \underline{b} = 1,2,\cdots, 8$
in the octet representation, and (\S \ref{sec:SL3C}) $\alpha, \underline{\beta} = 0,1,2,\cdots, 8$ in the reducible adjoint plus singlet representation (see below, and \S \ref{sec:SL3C} for component notation for a single qutrit).} ${R}^{a\underline{b}}$. 

We now apply the method of group characters for the enumeration of local unitary invariants at low degree in this case. The required notation and basic results from theory are set out in appendix \ref{sec:Schurology} which we briefly summarize here (see also \cite{jarvis:sumner:2012aith}). Returning to the coordinate system for the density matrix based on the fundamental representation, $\rho^{i\underline{j}, k\underline{\ell}}$ transforms as a direct product of the $9$-dimensional adjoint plus singlet representations on each subsystem, with group characters represented as $\{1\}\{\overline{1}\}\!\cdot\!\{1\}\{\overline{1}\}$.  The count of linearly independent polynomial invariants at each degree $n$ is then given by the (square of) the number of one dimensional representations in the \emph{plethysm} of characters,
$(\{1\}\{\overline{1}\}\!\cdot\!\{1\}\{\overline{1}\})\underline{\otimes}\{n\}$. This can be evaluated using the group theory package {\small \texttt{SCHUR} }\normalsize  \cite{SCHURsfg} as detailed in appendix \ref{sec:Schurology}. The resulting first few terms in the Molien series are (from Theorem 1, appendix \ref{sec:Schurology} and (\ref{eq:SU3SU3Molien}))
\begin{equation}
\label{eq:MolienLUqutrTerms}
h(z) = 1 + z + 4 z^2 + 11 z^3 + 34z^4 + 108z^5 + \cdots\,.
\end{equation}
Given that
\begin{equation}
\label{eq:MolienLUqutrRatl}
\frac{1+ \cdots}{(1-z)(1-z^2)^3(1-z^3)^7(1-z^4)^{17}\cdots} = 1 + z + 4 z^2 + 11 z^3 + 34z^4 +  \cdots\,,
\end{equation}
we infer the existence of a single linear invariant (the trace), as well as three quadratic, seven cubic and seventeen algebraically independent quartic invariants\footnote{In the absence of a complete evaluation for $h(z)$, it is not possible to confirm whether all of these invariants are indeed polynomially independent, as assumed in this trial generating function by setting the denominator to $1$ (see \cite{KingWelshJarvis2007}).}.

It is of some interest to identify a concrete set of invariants at each degree. In view of the importance of the octet basis for later sections, we proceed using the $r$, $\overline{r}$, $R$ components (in an obvious notation). Candidate invariants are found in principle by constructing all possible sums of words in the alphabet 
\[
\{ r^a, r^{\underline{a}}, R^{a\underline{a}}, \delta_{ab}, \delta_{\underline{a}\underline{b}}, 
f_{abc}, d_{abc}, f_{\underline{a}\underline{b}\underline{c}}, d_{\underline{a}\underline{b}\underline{c}};
a,b,c, \underline{a},\underline{b},\underline{c} = 1,\cdots,8  \}
\]
over which complete tensor contractions have been applied, and which are connected (that is, cannot be written as a product of two such totally contracted objects). Note that the polynomial degree (in $r$, $\overline{r}$ and $R$) of such an object does not include the count of invariant tensors $f$, $d$ and $\delta$ (the latter usually being suppressed in explicit constructions via the Einstein convention).

Table \ref{tab:QuadCub} gives a list of quadratic and cubic polynomials which account for the required number of independent invariants at these degrees. In the quartic case, there exists a large number of possibilities
for tensor contractions, from which the correct number of 17 linearly independent quantities must be identified.
Here (and as shown already in Table \ref{tab:QuadCub}) it is useful to adopt a more refined grading by degree, for invariant quantities $K_{pqs}$ of the form $r^p\overline{r}^qR^s$, and to adapt the group-theoretical character methods for counting invariants accordingly. This task is carried out in appendix \ref{sec:Schurology} and the results are summarized in Table \ref{tab:GradedCount}.
For combinatorial enumeration, at this degree it is slightly easier to use the original basis for the components of $\rho$ in the defining representation, and to transfer to the octet basis after the counting is established. For this purpose, an encyclopaedic source of identities and interrelations between the $f$ and $d$ coefficients is the paper \cite{macfarlane1968gell}; see also \cite{azcarraga:macfarlane:mountain:perezbueno:1998invariant}. The method is illustrated in appendix \ref{sec:Schurology} with the explicit construction of linearly independent candidates for the $K_{103}$, ${r}R^3$, and $K_{013}$, $\overline{r}R^3$, invariants, in both the defining and the octet basis, as well as explicit expressions for candidates for the five $K_{004}$, $R^4$ invariants in the defining representation.

In summary, it should be noted that the present methods for identifying LU invariants, based on character theory, are complementary to enumerative constructions of invariants via graphical techniques (deriving from trace and contraction operations on the tensor coordinates describing the state); 
see for example \cite{hero2009measProc,
hero2009stable,vrana2011local,vrana2011algebra,Szalay2012deg6lu}. For qu$\!\!D\!$it systems with small $D$, such as the bipartite mixed qubit case treated in our paper \cite{KingWelshJarvis2007}, and the qutrit case developed here, the group character methods do indeed give the direct count of all linearly independent, and algebraically independent, invariants at each degree (while Molien's theorem gives a closed form, which must be evaluated by integration over the group using the Haar measure, for the complete Molien series).

{\small \begin{table}
  \centering 
\begin{center}
  \begin{tabular}[tbp]{|l|l|}
\hline   
& \\
$K_{000}$ &$1$\\  
& \\
\hline
\end{tabular}
\hskip 3ex
  \begin{tabular}[tbp]{|l|l|}
\hline   
& \\
$K_{200}$ &$ r^a r^a$\\  
& \\
\hline
& \\
$K_{020}$ &$ r^{\underline{a}} r^{\underline{a}}$\\  
& \\
\hline
& \\
$K_{002}$ &$ R^{a\underline{b}} R^{a\underline{b}}$\\  
& \\
\hline
\end{tabular}
\hskip 3ex
  \begin{tabular}[tbp]{|l|l|}
\hline   
& \\
$K_{300}$ &$d_{abc}r^{a}r^{b}r^{c}$\\  
& \\
\hline
& \\
$K_{030}$ &$d_{\underline{a}\underline{b}\underline{c}}r^{\underline{a}}r^{\underline{b}}r^{\underline{c}}$\\  
& \\
\hline
& \\
$K_{111}$ &$r^{a}R^{a\underline{b}}r^{\underline{b}}$\\  
& \\
\hline
& \\
$K_{102}$ &$d_{\underline{a}\underline{b}\underline{c}}r^aR^{b\underline{d}} R^{c\underline{d}}$\\  
& \\
\hline
& \\
$K_{012}$ &$d_{\underline{a}\underline{b}\underline{c}}r^{\underline{a}}R^{d\underline{b}} R^{d\underline{c}}$\\  
& \\
\hline
& \\
$K_{003}^d$ &$d_{abc}d_{\underline{a}\underline{b}\underline{c}}R^{a\underline{a}}
R^{b\underline{b}}R^{c\underline{c}}$\\  
& \\
\hline
& \\
$K_{003}^f$ &$f_{abc}f_{\underline{a}\underline{b}\underline{c}}R^{a\underline{a}}
R^{b\underline{b}}R^{c\underline{c}}$\\  
& \\
\hline
\end{tabular} \\
\vskip3ex
\end{center} 
\caption{\protect{\small
Qutrit local unitary invariants for degree 1 (left), 2 (centre) and 3 (right) constructed as totally contracted, connected tensors in the components of the density matrix in the octet basis, together with the invariant tensors $f$, $d$ (and the orthogonal metric $\delta$, not explicitly written in summing over repeated indices). The index notation adoped for the labels $K_{pqr}$ reflects grading with respect to the powers of the
$r$, $\overline{r}$ and $R$ components of the density matrix in this basis. }}
\label{tab:QuadCub}
\mbox{}\\
\end{table}
}
\normalsize
{\small \begin{table}
  \centering 
\begin{center}
  \begin{tabular}[tbp]{|c|c|c|c|c|c|c|c|c|c|c|c|}
\hline 
&&&&&&&&&&&\\
~$400$~&~$040$~&$103$&$013$&
$202$&$022$&$112$& $121$&$211$&$301$&$031$&~ $004$~\\
&&&&&&&&&&&\\
\hline 
&&&&&&&&&&&\\
~$r^4$~&~$\overline{r}^4~$&$rR^3$&$\overline{r}R^3$&
$r^2R^2$&$\overline{r}^2R^2$&$r\overline{r}R^2$& $r\overline{r}^2R$&$r^2\overline{r}R$&$r^3R$&$\overline{r}^3R$&~ $R^4$~\\
&&&&&&&&&&&\\
\hline
\hline
&&&&&&&&&&&\\
0&0&2&2&2&2&2&1&1&0&0&5\\
&&&&&&&&&&&\\
\hline
\end{tabular} \\
\vskip3ex
\end{center} 
\caption{\protect{\small
Counting qutrit local unitary invariants at quartic degree, graded with respect to the powers of the
$r$, $\overline{r}$ and $R$ components of the density matrix in the octet basis. See appendix \ref{sec:Schurology} for details of the construction of these 17 quartic mixed state invariants.
 }}
\label{tab:GradedCount}
\mbox{}\\
\end{table}}
\normalsize

\section{The qutrit SLOCC transformation group $SL(3,{\mathbb C})$: two qutrit mixed state invariants and entanglement monotones}
\label{sec:SL3C}
We now extend the analysis beyond local unitary transformations, to the study of invariants of the density operator under more general quantum operations associated with measurement. Assuming as usual that any multi-outcome measurement can be realized as a composition of elementary two-outcome operations, we treat the case of measurement $\{ E_1,E_2\}$ with $E_1{}^\dagger E_1 + E_2{}^\dagger E_2 = I$, and 
transformation
\begin{equation}
\label{eq:RhoTransf}
\rho \rightarrow \rho' := p_1 \rho'_1 + p_2 \rho'_2, \qquad \rho'_1= \frac{E_1 \rho E_1{}^\dagger}{p_1},
\quad \rho'_2= \frac{E_2 \rho E_2{}^\dagger}{p_2}
\end{equation} 
with $p_1=Tr(E_1 \rho E_1{}^\dagger)$, and  $p_2=Tr(E_2 \rho E_2{}^\dagger)$. Consider moreover the case of local transformations $E= A\otimes B$, or compositions of one-sided operators of the type $A\otimes I$ or
$I \otimes B$.
The actions $\rho \rightarrow (A\otimes I)\rho(A^\dagger\otimes I)$, $\rho \rightarrow (A\otimes I)\rho(A^\dagger\otimes I)$ are not trace preserving in general and so do not constitute valid quantum operations. However, for invertible maps 
it is nonetheless fruitful to regard the transformations on $\rho$ in a projective sense, up to scalar multiplication to recover the correct unit trace normalization. 

Consider first a single qutrit mixed system with density operator $\varrho$. In view of the above, we append to the Gell-Mann matrices the identity ${\mathbb I}_3$ and introduce a corresponding additional ninth component of the density operator. In the octet basis we have extended coordinates
\[
\varrho = \textsl{r}^0 {\mathbb I}_3 + \sum_{a=1}^8 \textsl{r}^a \lambda_a \,, 
\qquad \textsl{r}^0 = \textstyle{\frac 13} Tr(\varrho), \qquad 
\textsl{r}^a = \textstyle{\frac 12} Tr(\varrho \lambda_a).
\]
Under transformations 
$\varrho \rightarrow A \varrho A^\dagger$ with $A$ invertible and of unit determinant, $A\in SL(3,{\mathbb C})$, we have that $Det(\varrho)$ is invariant. This condition can be expressed as a constraint on the transformations of the coordinates via
\[
6Det(\varrho) = \varepsilon^{ijk}\varepsilon_{\ell mn} \varrho_{i}{}^\ell \varrho_{j}{}^m \varrho_{k}{}^n\,.
\]
This can be written in the octet basis with the help of the standard identity
\[
Det(\varrho) = Tr(\varrho)^3 + 2Tr(\varrho^3) - 3Tr(\varrho)Tr(\varrho^2)
\]
and using the $\lambda$-matrix conventions of \cite{macfarlane1968gell}, we define the $9\times 9 \times 9$ totally symmetric three-fold tensor $\widetilde{d}_{\alpha\beta\gamma}$ via
\begin{align}
\textstyle{\frac 14 }Det(\rho) :=& \, \widetilde{d}_{\alpha\beta\gamma}\textsl{r}^\alpha \textsl{r}^\beta \textsl{r}^\gamma
= \textstyle{\frac{3}{2}} (\textsl{r}^0)^3 -   \textstyle{\frac{3}{2}}\textsl{r}^0 \textsl{r}^a \textsl{r}^a + 
\textsl{r}^a \textsl{r}^b \textsl{r}^c d_{abc}\,,
\nonumber
\end{align}
(with repeated index summations $a,b,c=1,\cdots,8$ and $\alpha,\beta,\gamma=0,1,2,\cdots,8$) so that the nonzero entries are
\begin{equation}
\widetilde{d}_{000}=\textstyle{\frac{3}{2}},\quad \widetilde{d}_{00a} = \widetilde{d}_{0a0}=\widetilde{d}_{a00}=0, \quad \widetilde{d}_{0ab}=\widetilde{d}_{a0b}=\widetilde{d}_{ab0}= -\textstyle{\frac{1}{2}}\delta_{ab},
\quad \widetilde{d}_{abc}=d_{abc}\,.
\end{equation}

From the above data  we define $H_d(8,1)$ to be the subgroup of $GL(9,{\mathbb R})$ ($9\times 9$ invertible real matrices) which preserve the tensor $\widetilde{d}_{\alpha \beta\gamma}$, that is,
\begin{equation}
H_d(8,1) = \{ m \in GL(9,{\mathbb R}) : m_\alpha{}^{\alpha'}m_\beta{}^{\beta'}m_\gamma{}^{\gamma'}\widetilde{d}_{\alpha'\beta'\gamma'} = \widetilde{d}_{\alpha\beta\gamma} \}.
\end{equation}
Obviously $H_d(8,1)$ is a group, and since the above conditions are continuous in the standard metric, it is closed. Hence it is indeed a \emph{bona fide} matrix subgroup of $GL(9,{\mathbb R})$. Moreover, the linear mapping
induced by $A \varrho A^\dagger \equiv m_A{}^\alpha{}_\beta \varrho^\beta \lambda_\alpha $ for $A\in SL(3,{\mathbb C})$ provides a 3:1 homomorphism\footnote{$A$, $\omega A$, $\omega^2 A$ all produce the same action on $\varrho$ for $\omega^3=1$.}: $SL(3,{\mathbb C})\rightarrow H_d(8,1)$. 
We now proceed to show directly the isomorphism of the corresponding Lie algebras,  
$h_d(8,1)\cong sl(3,{\mathbb C})_{\mathbb R}$.

Identifying $h_d(8,1)$ as usual as the tangent space at ${\mathbb I}$, we have for $x_\alpha{}{}^\beta \in h_d(8,1)$ by differentiation
\[
x_\alpha{}{}^{\alpha'}\widetilde{d}_{\alpha'\beta \gamma} + 
x_\beta{}{}^{\beta'}\widetilde{d}_{\alpha\beta' \gamma}+x_\gamma{}{}^{\gamma'}\widetilde{d}_{\alpha\beta \gamma'} = 
0
\]
and thus examine the four different cases $\widetilde{d}_{\alpha \beta\gamma}$ as above. Defining $u_a{}^b=\textstyle{\frac{1}{2}}\big(x_a{}^b\!+\!x_a{}^b\big)$, 
$v_a{}^b=\textstyle{\frac{1}{2}}\big(x_a{}^b\!-\!x_a{}^b\big)$. We find:
\begin{align}
\widetilde{d}_{000}:\qquad x_0{}{}^{0}=& \, 0\,; \nonumber \\
\widetilde{d}_{00a} :\qquad  x_0{}{}^{a}=& \,\textstyle{\frac{3}{2}}x_a{}{}^{0}\,; \nonumber \\
 \widetilde{d}_{0ab}:\qquad 
 u_a{}{}^{b} = & \,\,x_0{}{}^{e}d_{eab} \,;   \nonumber \\
\widetilde{d}_{abc}:\quad
\textstyle{\frac{1}{2}}\big(x_a{}^0\delta_{bc} \!+\!x_a{}^0\delta_{bc}\!+\!x_a{}^0\delta_{bc}\big)
= &\, \big(u_a{}^e d_{ebc} \!+\! u_b{}^e d_{aec} \!+\!u_c{}^e d_{abe}\big) 
\!+\!\big(v_a{}^e d_{ebc} \!+\! v_b{}^e d_{aec} \!+\!v_c{}^e d_{abe}\big).
\nonumber
\end{align} 
Substituting third condition into the last, with the help of the second relation, yields the cyclically permuted once-contracted $(dd)$ quartet which by a standard identity (for more details see the related discussion in appendix \ref{sec:Schurology} below, and \cite{macfarlane1968gell}) precisely cancels the cyclic $(\delta\delta)$ form on the left hand side, leaving 
\[
v_a{}^e d_{ebc} \!+\! v_b{}^e d_{aec} \!+\!v_c{}^e d_{abe} =0.
\]
At the same time from antisymmetry, $v_a{}^b \in so(8)$, as it preserves the metric $\delta_{ab}$. It is well known that $su(3)$ is a maximal (simple) subalgebra of $so(8)$, indeed $v_a^b = f_{abc}$ solves this constraint by virtue of the cyclic $(df)$ quartet identity (for details see appendix \ref{sec:Schurology} below and \cite{macfarlane1968gell}). 

From the above, the (real) Lie algebra $h_d(8,1)$ is sixteen dimensional, spanned by the adjoint octet $ (F_a )_b{}^c=f_{abc}$, $ (F_a )_0{}^b= (F_a )_0{}^b=0$, together with the additional octet $ (D_a )_0{}^b=\delta_a{}^b$, $ (D_a )_b{}^0=\textstyle{\frac 23}\delta_a{}^b$, $ (D_a )_b{}^c=d_{abc}$. The  isomorphism 
$F_a\rightarrow \lambda_a, D_a\rightarrow i\lambda_a$ with $sl(3,{\mathbb C})_{\mathbb R}$ is verified with the help
of another $(dd)$ to $(f\!\!f)$ identity, this time of antisymmetric rather than cyclic type.

Clearly the qubit SLOCC group $SL(2,{\mathbb C})_{\mathbb R} \cong SO(3,1)$ (reviewed in appendix \ref{sec:TwoQubitReview} below) finds a precise parallel with $H_d(8,1)$ in
the above qutrit setting. We here identify $H_d(8,1)$ as the appropriate qutrit `relativistic' type symmetry group, and explore its role in two qutrit entanglement. As discussed in appendix \ref{sec:TwoQubitReview}, one way to proceed is simply to identify relevant polynomial invariant quantities, once again using group character methods adapted from the $SO(3,1)$ case to this case. In previous work \cite{FauserJarvisKing2006nbr} we have investigated extensions of group character methods (as explained above in the enumeraton of local unitary invariants) to `non-classical' matrix groups, and  $H_d(8,1)$ is such a group (although it happens to be locally isomorphic to
$SL(3,{\mathbb C})_{\mathbb R}$ in this case). The relevant theory for computing the Molien series is outlined in  appendix \ref{sec:Schurology} (see Theorem 2 and equation \ref{eq:MolienHd81app}), and the result is conjectured to be
\begin{equation}
\label{eq:MolienHd81}
1+z^3+2 z^6+5 z^9+12 z^{12} +\cdots = \frac{1+ \cdots}{(1-z^3)(1-z^6)(1-z^9)^3(1-z^{12})^6\cdots}\, ,
\end{equation}
suggesting the existence of a single invariant at each of degrees 3 and 6, three at degree 9, and six at degree 12 in this case.

To exemplify the method, we give the construction for the lowest degree quantities. Recall that the standard coordinates of the density operator are given as $r^a$, 
$r^{\underline{a}}$, $R^{a\underline{a}}$, or simply $r^{\alpha \underline{\alpha}}$ in the $9\times 9$ basis. By analogy with the method applied in the local unitary case, LSL invariants are now connected, totally contracted words in $\{r^{\alpha \underline{\alpha}}, \widetilde{d}_{\alpha\beta\gamma}, 
\widetilde{d}_{\underline{\alpha}\underline{\beta}\underline{\gamma}} \}$ written with summation over repeated indices.

At degree three, there is clearly only one invariant,
\begin{equation}
C_3 = \widetilde{d}_{\alpha\beta\gamma}\widetilde{d}_{\underline{\alpha}\underline{\beta}\underline{\gamma}}
r^{\alpha \underline{\alpha}}r^{\beta \underline{\beta}}r^{\gamma \underline{\gamma}}\,.
\end{equation} 
For degree 6 we must examine the different possible index connectivities. 
Without loss of generality, we can order the six $r$ tensors so that their first (non-underlined) index follows the ordering of the corresponding $\widetilde{d}$ subscripts. There remain in principle $\texttt{C}^6_3=20$ possible choices of index matchings for the remaining underlined indices, to partners on their respective $\widetilde{d}$ tensors. However, the total symmetry, and equivalence of, the two $\widetilde{d}$ tensors, and index re-labellings, lead to only one independent form which can be taken as
\begin{align}
C_6 = & \, \widetilde{d}_{\alpha\beta\gamma}\widetilde{d}_{\rho\sigma\tau}
r^{\alpha \underline{\alpha}}r^{\beta \underline{\beta}}r^{\gamma \underline{\gamma}}
r^{\rho \underline{\rho}}r^{\sigma \underline{\sigma}}r^{\tau \underline{\tau}}
\widetilde{d}_{\underline{\beta}\underline{\gamma}\underline{\rho}}
\widetilde{d}_{\underline{\sigma}\underline{\tau}\underline{\alpha}}\, , 
\end{align}
which exemplifies the pattern of cross-overs in the matchings between the underlined $r$ indices, and their corresponding $d$ subscripts.

To end our analysis, we give the explicit form of the $H_d(8,1)$ cubic invariant $C_3$ in terms of the previously-identified local unitary invariants of Table \ref{tab:QuadCub} (see \S \ref{sec:LocalUnitary} above). For trace-normalized density operators, $C_3$ will of course no longer be a homogeneous quantity.
Using the transcription between the components $r^{\alpha \underline{\alpha}}$ and the octet basis $\{r, \overline{r}, R\}$, and using the above identification of the components of the extended $\widetilde{d}$ tensors, we find 
\begin{equation}
C_3=K^d_{003}+\textstyle{\frac 32}\big(K_{300}+K_{030}\big)+\textstyle{\frac 32}\big(K_{111}-K_{102}-K_{012}\big)
-\textstyle{\frac 14}\big(K_{200}+K_{020} \big)+\textstyle{\frac{1}{12}}K_{002}+\textstyle{\frac{1}{324}}\,.
\end{equation}

The interpretation of such invariants to characterise entanglement is discussed in appendix 
\ref{sec:Monotones}. In particular, it is proven how appropriate powers of (the absolute value of) such homogeneous LSL polynomial invariants lead to \emph{bona fide} entanglement monotones. In the present case, the analysis shows that the quantity $|C_3|^{\frac 13}$ will be such a quantity. Similarly, at degree 6 the quantity $|C_6|^{\frac 16}$ will provide a further example.

\section{Conclusions.}
\label{sec:Conc}
This paper has developed a group-theoretical analysis of the density operator for the two qutrit system, and its local invariants. Starting with the group $SU(3)\times SU(3)$ of local unitary transformations, character methods have been used to enumerate and construct polynomial invariants (up to degree four) in the components of the density operator (\S \ref{sec:LocalUnitary}). A larger, SLOCC type  transformation group acts on the density operator regarded projectively (up to normalization to unit trace). This is the group $SL(3,{\mathbb C})_{\mathbb R}$, presented as a group of real $9\times 9$ matrices (a matrix subgroup of $GL(9,{\mathbb R})$) acting linearly on the 9-dimensional space of projective coordinates for the qutrit density matrix.
This group (which is denoted $H_d(8,1)$ in this context), has been explicitly identified in \S \ref{sec:SL3C}, and its character theory discussed, in order to identify appropriate LSL polynomial invariants.
The details of the group character methods are outlined in the appendix to the paper, which also includes a review of 
our earlier paper \cite{KingWelshJarvis2007} on the two qubit system and its ring of local unitary invariants. A further appendix discusses how polynomial invariants can be used to form entanglement monotones, and exemplifies this for the two qubit case for some standard $SL(2,{\mathbb C})$ invariants. In the same vein, in \S \ref{sec:SL3C} we count and construct candidates for independent two qutrit entanglement monotones, giving in particular the explicit form for the lowest degree, cubic invariant in terms of the linear, cubic and quadratic local unitary invariants identified in \S \ref{sec:LocalUnitary}.
\vfill
\noindent
\textbf{Acknowledgements}\\
PDJ thanks S Szalay (Wigner Research Centre, Budapest) for fruitful correspondence, for pointing out key literature on entanglement monotones, and for generously providing his own unpublished notes (see also \cite{szalay2013quantum}). Discussion and correspondence on aspects of this work with
G Barwick, A Bracken, D Ellinas, B Fauser, S Jacobsen, R King, P Levay, A Ratcliffe, J Sumner and T Welsh are gratefully acknowledged.

\newpage
\begin{appendix}
\renewcommand{\theequation}{\thesection-\arabic{equation}}
\section{Counting invariants for mixed state systems via group characters.}
\label{sec:Schurology}
\subsection{LU invariants for qubits and qutrits}
In this section, for the sake of completeness, we give a brief outline of the combinatorial description of the representations and characters of symmetry groups arising as transformations on the density operator for bipartite qubit and qutrit mixed systems. The enumeration of polynomial invariants (at low degree) will then become possible, with a knowledge of standard manipulations of the required characters (as mentioned in \S \ref{sec:Intro}, this method is not in general sufficient to derive the full structure of the Molien series). 

We refer to \cite{KingWelshJarvis2007} for details of the two qubit case, and to \cite{jarvis:sumner:2012aith} for a general review from which the following is adapted. The mathematical setting for the study of entanglement invariants for $K$-fold composite qu$\!\!D\!$it systems, is that there is a model space $V$ which is a $K$-fold tensor product, $V \cong {\mathbb C}^D\otimes {\mathbb C}^D\otimes \cdots 
\otimes  {\mathbb C}^D$. The components of $V$ in some standard basis describe the state; for example in Dirac notation a pure state is a ket $|\psi \rangle \in V$ of the form
$
|\psi \rangle = \sum_{1}^{D} \psi^{i_1 i_2 \cdots i_K} |i_1,i_2, \cdots, i_K \rangle
$
in the case of qu$\!\!D\!$its. For mixed states, the model space is instead $W =V \otimes V^*$ (a $K$-fold tensor product of ${\mathbb C}^D\otimes {\mathbb C}^D{}^*$), and coordinatized via the density operator $\rho \in W$. For present purposes we illustrate only the bipartite case,
\begin{equation}
\rho =\sum_{1}^{D} \rho^{k\underline{\ell};i\underline{j}}|i,\underline{j}\rangle \langle k ,\underline{\ell}|
\end{equation}
where we have introduced the underline index convention for denoting components with respect to the second subspace of the composite system. 

We focus attention on the linear action of the appropriate matrix group $G = G_1 \times G_2 \times \cdots \times G_K$ on $V$ and $W$. In the qu$\!\!D\!$it case each local group $G_k$ is a copy of $U(D)$, but given the irreducibility of the fundamental representation, for polynomial representations, the analysis can be done using the character theory of the complex group. We compute the Molien series $h(z) = \sum_0^\infty h_n z^n$ degree-by-degree using combinatorial methods based on classical character theory \cite{Weyl1939,littlewood1940}. Characters of $GL(D)$ 
and $SL(D)$ differ only by powers of the determinant character (and similarly for $U(D)$ and $SU(D)$). All evaluations are carried out using the group representation package {\small \texttt{SCHUR} }\normalsize \cite{SCHURsfg}. 

In terms of class parameters
(eigenvalues) $x_1,x_2,\cdots, x_D$ for a nonsingular matrix $m \in GL(D)$, the defining representation, the character is simply $Tr(m) =  x_1+ x_2+ \cdots + x_D$; the contragredient has character
$Tr(m^T{}^{-1}) =  x_1{}^{-1}+ x_2{}^{-1}+ \cdots + x_D{}^{-1}$. Irreducible polynomial and rational characters of $GL(D)$ are given in terms of the celebrated Schur functions \cite{Weyl1939,littlewood1940} denoted $s_\lambda(x)$, where $\lambda = (\lambda_1,\lambda_2,\cdots,\lambda_D)$, $\lambda_1 \ge \lambda_2 \ge \cdots \ge \lambda_D$, is an integer partition of at most $D$ nonzero parts. 
$\ell(\lambda)$, the length of the partition, is the index of the last nonzero entry (thus $\ell(\lambda)=D$ if $\lambda_D >0$). $|\lambda|$, the weight of the partition, is the sum $|\lambda|=\lambda_1+\lambda_2 + \cdots + \lambda_D$, and we write $\lambda \vdash |\lambda|$. For brevity we write the Schur or $S$-function simply as $\{\lambda \}$ where the class parameters are understood\footnote{Partitions $\lambda$ are also abbreviated as words in monoid style, thus $(2^3) \equiv (2,2,2)$, etc.}. Thus the space $V$ as a representation of $G$ as a $K$-fold Cartesian product is endowed with the corresponding product of $K$ characters of the above defining representation of each local group, $\chi= \{1\} \cdot \{1\}\cdot \, \cdots \,  \cdot \{1\}$ in the quantum mechanical pure state case, and 
$\chi = (\{1\} \{\overline{1}\}) \! \cdot \! (\{1\} \{\overline{1}\})\!\cdot \, \cdots \, \cdot\!(\{1\} \{\overline{1}\})$ in the quantum mechanical mixed state case, where $\{1\}$ is the character of the defining representation, and $\{\overline{1}\}$ that of its contragredient.
The space of polynomials of degree $n$ in $\psi$ or $\rho$ is a natural object of interest, and by a standard result is isomorphic to the $n$-fold symmetrised tensor product of $n$ copies of $V$ or $W$. Its character is determined by the corresponding Schur function \emph{plethysm}, 
$\chi \underline{\otimes} \{n\}$, and the task at hand is to enumerate the one-dimensional representations occurring therein.

Before giving the relevant results it is necessary to note two further rules for combining Schur functions. The \emph{outer} Schur function product, is simply the pointwise product of Schur functions, arising from the character of a tensor product of two representations. Of importance here is the \emph{inner} Schur function product $\ast$ defined via the Frobenius mapping between Schur functions and irreducible characters of the symmetric group. We provide here only the definitions sufficient to state the required counting theorems in technical detail. For a Hopf-algebraic setting for symmetric functions and characters of classical (and some non-classical) groups see also \cite{FauserJarvis2003hl,FauserJarvisKing2006nbr,FauserJarvisKing2013has}. 

Concretely, we introduce outer (pointwise) products in the Schur function basis as follows:
\[
\{\lambda \}  \{ \mu \} = \sum_\nu C^\nu_{\lambda,\mu} \{\nu \},
\]
where the $C^\nu_{\lambda,\mu}$ are the famous `Littlewood-Richardson' coefficients. Closely related is the dual operation of skew, which we note here for completeness:
\[
\{\lambda \} / \{ \mu \} = \sum_\nu C^\lambda_{\mu,\nu} \{\nu \}.
\]
Similarly, we introduce structure constants for inner products in the Schur function basis:
\[
\{\lambda \} \ast \{ \mu \} = \sum_\nu g^\nu_{\lambda,\mu} \{\nu \}.
\]
For partitions $\lambda$, $\mu$ of equal weight\footnote{If 
$|\lambda| \ne |\mu|$ then $\{\lambda \} \ast \{ \mu \}=0$.}, $|\lambda| = |\mu|= n$, say, this expresses the reduction of a tensor product
of two representations of the symmetric group ${\mathfrak S}_n$ labelled by partitions $\lambda$, $\mu$. By associativity, we can extend the definition of the structure constants to $K$-fold inner products,
\[
\{\tau_1 \} \ast \{\tau_2 \} \ast \cdots \ast \{\tau_K \} = \sum_\nu g^\nu_{\tau_1, \tau_2, \cdots, \tau_K}
\{\nu \}.
\] 

\noindent
\textbf{Theorem 1: Counting unitary invariants \cite{jarvis:sumner:2012aith}}
\begin{description}
\item[(a) Pure states, $K$-fold composite system]\mbox{}\\
Let $D$ divide $n$, $n = rD$, and let $\tau$ be the partition $(r^D)$ (that is, with Ferrers diagram a rectangular array of $r$ columns of length $D$). Then the number $h_n$ of linearly independent invariants is
\[
h_n = g^{(n)}_{\tau,\tau,\cdots,\tau}\quad \mbox{($K$-fold inner product)}.
\] 
If $D$ does not divide $n$, then $h_n =0$.
\item[(b) Mixed states, bipartite system \cite{KingWelshJarvis2007}]\mbox{}\\ The number $h_n$ of linearly independent invariants is
\[
h_n = \sum_{|\tau|= n,\ell(\tau) \le D^2}
\left( \sum_{|\sigma|= n, \ell(\sigma) \le D} g^{\tau}_{\sigma,\sigma}\right)^{\!\!\!\!2}.
\]
\mbox{}\hfill $\Box$
\end{description}
\subsection*{$SU(2)\times SU(2)$ invariants for 2 qubit mixed systems}
Application of the above counting theorem for the two qubit mixed system was carried out in \cite{KingWelshJarvis2007}. For details we refer to appendix \ref{sec:TwoQubitReview} below, which gives a review of this work and its extension to constructions of entanglement monotones (the Molien series is given in equation (\ref{eq:MolienSU2SU2})).
\subsection*{$SU(3)\times SU(3)$ invariants for 2 qutrit mixed systems}
Counting the first few coefficients $h_n$ using the above theorem yields the Molien series for the two qutrit mixed system,
\begin{equation}
\label{eq:SU3SU3Molien}
h(z) = 1 + z + 4 z^2 + 11 z^3 + 34z^4 + 108z^5+ \cdots\,.
\end{equation}
Given that
\begin{equation}
\label{eq:SU3SU3MolienRatl}
\frac{1+ \cdots}{(1-z)(1-z^2)^3(1-z^3)^7(1-z^4)^{17}\cdots} = 1 + z + 4 z^2 + 11 z^3 + 34z^4 + \cdots\,,
\end{equation}
we look to construct (apart from the trace), three quadratic, seven cubic and seventeen algebraically independent quartic invariants (see \S \ref{sec:LocalUnitary} above).  On the basis of these partial results, the count of invariants and the corresponding invariant ring are considerably richer than those for the two qubit mixed system\footnote{Establishing the full generating function via Molien's theorem would require using the Haar measure on the $SU(3)\times SU(3)$ group.}.
Candidates for the quadratic and cubic invariants are listed in Table \ref{tab:QuadCub} above. For the seventeen quartic invariants, a count of contributions at separate degrees in each of the contributing tensors $r$, $\overline{r}$, $R$ aids identification (see Table \ref{tab:GradedCount} above). 

In order to arrive at these assignments, the group character arguments can be adapted as follows. Any invariant quantity $K_{pqs}$ at degree $r^p\overline{r}^qR^s$ must arise as an admissible coupling (to the trivial representation) between all irreducible representations occurring in the respective symmetrized products. Thus individual group character plethysms of the respective powers $p,q,s$ must be derived, and the number of invariants is simply the coefficient of the trivial character after taking the outer product\footnote{This problem also arises in the context of constructing non-subgroup labelling operators for a certain group embedding, in this case $SU(9)$ in a basis of $SU(3)\times SU(3)$. The transcription to the equivalent embedding for the qubit case has been discussed in 
\cite{KingWelshJarvis2007} and the corresponding $SU(4)$ labelling problem treated in \cite{Quesne1976:sea}. }. Working with $S$-function notation for $SU(3)$, and (octet) adjoint representation $\{21\}$, we have for the direct product characters
\[
r \cong \{21\} \cdot 1, \qquad \overline{r}\cong 1\cdot \{21\}, \qquad R \cong \{21\}\cdot \{21\}
\]
where for better readability the trivial character has here been written simply as `$1$' (it appears formally as $\{0\}$ below) . Thus the number $H_{pqs}$ of such invariants (giving the generating function $H(x,y,z) = \sum H_{pqs}x^py^qz^s$ with $h(z) = H(z,z,z)$) is
\[
H_{pqs} = \left(\big(\{21\} \cdot 1\big)\underline{\otimes}\{p\}\right)
\left(\big(1\cdot \{21\}\big)\underline{\otimes}\{q\}\right) \left(\big(\{21\} \cdot \{21\}\big)\underline{\otimes}\{s\}\right)
\bigg|_{\{0\}\cdot\{0\}}
\]

Using this result we can enumerate the required contributions term by term from the list of contributing powers,
namely $301, 031, 103, 013, 202, 022, 112, 121, 211$, and $004$, as follows. All character manipulations are performed with {\small \texttt{SCHUR} }\normalsize \cite{SCHURsfg} for the group $SU(3)$.
\begin{description}
\item[301, 031]\mbox{}\\
We have 
\[
\{21\}\underline{\otimes}\{3\} = \{63\} + \{42\} + \{3 \} +  \{3^2\} + \{21\} + \{0\}\,.
\]
The $\{3\}$ plethysm of one octet is accompanied by a singlet (no involvement of the other octet) and so remains one-sided, $\{21\}\underline{\otimes}\{3\}\cdot \{0\}$ or
its left-right swapped version. In the final outer product with the tensor $\{21\}\cdot \{21\}$, there will be on one side a left over $\{21\}\{0\}=\{21\}$, which remains an octet. Thus there is no contribution at this grading.
\item[103,013]\mbox{}\\
In this case the $\{3\}$ plethysm applies to the tensor direct product $\{21\} \cdot \{21\}$, and by a standard distributivity property \cite{fauser:jarvis:king:2010a:vertex}, devolves to the evaluation of the sum of direct products of plethysms for all partition types on each side. The total count for each grading $103,013$ thus requires accumulating all resulting terms of the form $\{21\}\cdot \{0\}$, $\{0\}\cdot \{21\}$ respectively, which can couple to the remaining octet:
\begin{align}
\{21\}\underline{\otimes}\{3\} = & \,\{63\} + \{42\} + \{3 \} +  \{3^2\} + {\mathbf \{21\} }+ {\mathbf\{0\} }\,, \nonumber \\
\{21\}\underline{\otimes}\{21\} = & \, \{51\} + \{54\} + 2\{42\} + \{3\} +  \{3^2\} + 3\{21\}\,, \nonumber \\
\{21\}\underline{\otimes}\{1^3\} = & \,\{42\} + \{3 \} +  \{33\} + {\mathbf\{21\} } + {\mathbf\{0 \} }\,, \nonumber 
\end{align}
in which the first and third lines, $\{21\}\underline{\otimes}\{3\}\cdot \{21\}\underline{\otimes}\{3\}$ and
$\{21\}\underline{\otimes}\{1^3\}\cdot\{21\}\underline{\otimes}\{1^3\}$, will each contribute 1 such term.
\item[202, 022]\mbox{}\\
We compute
\begin{align}
\{21\}\underline{\otimes}\{2\} = & \, \{42\} + \{21\} + \{0\}\,, \nonumber \\
\{21\}\underline{\otimes}\{1^2\} = & \, \{3\} + \{3^2 \} + \{21\} \,. \nonumber 
\end{align}
These are to be applied to $\{21\} \cdot \{21\}$, which is to couple only to the symmetric $\{2\}$ plethysm of one of the other octets. Thus only the $\{2\}$ plethysm itself can provide the required singlet on one side. On the other hand, the option of having both quadratic terms coupling to $\{0\}$ would simply amount to a disconnected term. Thus, the only couplings to an overall singlet arise from outer products either between the two resulting octets, $\{21\}\{21\}$, or the two 27's, $\{42\}\{42\}$, and we infer 2 contributions for each of these gradings.
\item[112]\mbox{}\\
Both symmetrization and antisymmetization of the tensor $\{21\}\cdot\{21\}$ again include $\{21\}\cdot\{21\}$, which on each side can couple to the two octets, again giving 2 contributions. 
\item[121, 211]\mbox{}\\
Again the symmetrization entailed in each quadratic term can only couple to an invariant through octet $\{21\}$ contributions, and there is thus just 1 further invariant in each case.
\end{description}

Discounting the $400$ and $040$ gradings (which cannot yield totally connected tensor contractions) leaves a deficit of 5 invariants out of the required 17 (compare Table \ref{tab:GradedCount}) to come from the remaining case $004$, that is, pure quartic terms in $R$. The counting can indeed be confirmed by examining all
$\{21\}\underline{\otimes}\{\sigma\}$ character plethysms for all partitions of weight 4, $\sigma \vdash 4$, in the same manner as with the gradings treated above\footnote{The result is 6 linearly independent terms, which includes 
the disconnected form $(RR)^2= K_{002}^2$ (see \S \ref{sec:LocalUnitary} above).}. However, to show the connections with concrete instances of the claimed invariants, we provide here instead, a combinatorial argument to support the existence of the 5 remaining invariants at this degree.

Since the argument will involve index matchings based on the components of the density operator in the original, defining representation, we first examine a simpler case where the count is already established, and we confirm the correct number of terms by explicit computation. This is done below for the two distinct contributions identified above for the $103$ grading (the $013$ case is analogous). Having established how tensor contractions can be transcribed in principle into the octet basis, we then present an argument for the required five $004$-type quartic terms.

Consider then possible total tensor contractions for terms of the form $rRRR$. In the defining representation, 
$r$ has components $r^i{}_j$, and $R$ has components $R^{i\underline{p}}{}_{j \underline{q}}$ (these must in fact be traceless tensors). Without loss of generality, the total contraction of the underlined indices on the three $R$ tensors can be arranged in cyclic order, with the contraction of the non-underlined indices with the additional $r^i{}_j$ still to be applied:
\[
r^i{}_jR^{\cdot \underline{p}}{}_{\cdot \underline{q}}R^{\cdot \underline{q}}{}_{\cdot \underline{r}}R^{\cdot \underline{r}}{}_{\cdot \underline{p}} \,.
\]
The possibilities for the remaining contractions of $i,j$ can be enumerated by choosing two positions from the three $R$ factors, giving 6 options $(1,2)$, $(2,1)$, $(1,3)$, $(3,1)$, $(2,3)$, $(3,2)$; however, cyclically reordering the factors reveals only two distinct forms (avoiding traces), represented by
\[
(1,2) = r^i{}_jR^{k \underline{p}}{}_{i \underline{q}}R^{j \underline{q}}{}_{\ell \underline{r}}R^{\ell \underline{r}}{}_{k \underline{p}} \,,\qquad 
(2,1) = r^i{}_jR^{j \underline{p}}{}_{\ell \underline{q}}R^{k \underline{q}}{}_{i \underline{r}}R^{\ell \underline{r}}{}_{k \underline{p}} \, .
\]
Now assuming that (up to normalization) the transcription between adjoint indices (traceless matrices) and the octet basis follows $r^i{}_{j} \leftrightarrow r^d\big(\lambda_d\big){}^i{}_j$, we can match the above index orderings to traces of Gell-Mann matrices from each subspace:
\[
(1,2) \rightarrow Tr\big(\lambda_{\underline{a}} \lambda_{\underline{b}}\lambda_{\underline{c}}\big)R^{a \underline{a}} R^{b \underline{b}}R^{c \underline{c}}r^d Tr\big(\lambda_d \lambda_b \lambda_c \lambda_a\big),\,
(2,1) \rightarrow Tr\big(\lambda_{\underline{a}} \lambda_{\underline{b}}\lambda_{\underline{c}})R^{a \underline{a}} R^{b \underline{b}}R^{c \underline{c}}r^d Tr\big(\lambda_d \lambda_a \lambda_c \lambda_b\big)\,.
\]
Using standard trace identities shows that certain terms cancel due to symmetry incompatibilities between $f$ and $d$ tensors, and for $(1,2)$ there remain the index patterns (up to proportionality)
\[
d_{\underline{a}\underline{b}\underline{c}}\delta_{ac}\delta_{bd}\,, \,\,
d_{\underline{a}\underline{b}\underline{c}} (dd)_{ac,bd}\,, \,\,
d_{\underline{a}\underline{b}\underline{c}} (df)_{ac,bd}\,,\,\,
f_{\underline{a}\underline{b}\underline{c}} (f\!\!d)_{ac,bd}\,, \,\, \mbox{and}\,\,
f_{\underline{a}\underline{b}\underline{c}} (f\!\!f)_{ac,bd} \,, 
\]
where the quartets are once-contracted forms, for example $(dd)_{ac,bd} := d_{ace}d_{ebd}$, and so on.
By re-labelling, in the presence of $R^{a \underline{a}} R^{b \underline{b}}R^{c \underline{c}}$, each of these can be replaced by its cyclic sum (over $a,b,c$). Under such cyclic sums, standard identities
(see \cite{macfarlane1968gell,azcarraga:macfarlane:mountain:perezbueno:1998invariant}) show that the $(df)$ and $(f\!\!f)$ quartets vanish, while the $(dd)$ quartet reduces to the first, $(\delta\delta)$ term. The only remaining couplings are thus the first and third, $(\delta\delta)$ and $(f\!\!d)$ contributions, which moreover differ in sign between $(1,2)$ and $(2,1)$ above. By taking appropriate linear combinations we can separate these, and finally we identify the required 2 independent candidates for ${103}$ graded invariants in the octet basis as
\begin{align}
K_{103} = & \,d_{\underline{a}\underline{b}\underline{c}}
R^{a \underline{a}} R^{a \underline{b}}R^{d \underline{c}}r^d \, \nonumber \\
K'_{103} = & \, f_{\underline{a}\underline{b}\underline{c}}
R^{a \underline{a}} R^{b \underline{b}}R^{c \underline{c}}r^d(f\!\!d)_{ab,cd}\,.
\nonumber
\end{align}
Correspondingly, there will also be two $013$ graded invariants $K_{013}, K_{013}'$, defined by interchanging the roles of underlined and non-underlined indices.

Turning now to the $004$ graded ($R^4$) invariants, we provide the following combinatorial argument based on the defining representation, and omit the subsequent transcription to the octet basis with $f$ and $d$ coupled tensors, which can be carried out as done above for grading $103$. As with the $103$ case, we start with\footnote{For completeness we should also consider the case of a product of two separate traces in the underlined
indices, $R^{i \underline{p}}{}_{j \underline{q}}R^{\cdot \underline{q}}{}_{\cdot \underline{p}}R^{\cdot \underline{r}}{}_{\cdot \underline{s}}R^{\cdot \underline{s}}{}_{\cdot \underline{r}}$. The `diagonal' trace option (2,2) for the non-underlined indices will amount to the afore-mentioned disconnected form. In the octet basis for the underlined indices, the resulting $(\delta\delta)$ type coupling will in any case arise for the remaining options, as a part of the couplings generated from the totally connected underlined trace form treated in the text. For example, under cyclic symmetry, it is linearly related to cyclic sums of $(dd)$ quartets. An analogous comment applies for the
$(1,1)$, $(2,2)$ and $(3,3)$ type options in the $103$ and $013$ cases.} 
\[
R^{i \underline{p}}{}_{j \underline{q}}R^{\cdot \underline{q}}{}_{\cdot \underline{r}}R^{\cdot \underline{r}}{}_{\cdot \underline{s}}R^{\cdot \underline{s}}{}_{\cdot \underline{p}} \,.
\]
and consider the patterns of subsequent tensor contractions with the remaining unassigned indices. These can be enumerated by citing the positions of the two $R$ terms at which the ${}^i{}_j$ superscript and subscript of the first $R$ term are summed; the remaining contractions are then determined automatically. For example the $(4,2)$ case would be
\[
(4,2)=R^{i \underline{p}}{}_{j \underline{q}}R^{j \underline{q}}{}_{k \underline{r}}R^{k \underline{r}}{}_{\ell \underline{s}}R^{\ell \underline{s}}{}_{i \underline{p}}\,.
\]
There are thus 9 cases, but again these are subject to re-arrangement and cyclic re-ordering moves, under which it is easy to see that there are only five distinct terms:
\begin{align}
  (3,3); \quad (2,\,&4); \quad (4,2); \nonumber \\
(2,2) \leftrightarrow &\,\, (4,4);  \nonumber \\
(3,2) \leftrightarrow  (2,3) \leftrightarrow  &\, \,(3,4) \leftrightarrow  (4,3)\,.
 \nonumber 
\end{align}

We tentatively suggest then, that the required five $K_{004}$ type invariants can identified with these five distinct terms. With this discussion we complete the identification of the full complement of seventeen algebraically independent quartic invariants for the two qutrit mixed system.
\subsection{LSL invariants for qubits and qutrits}
As discussed in \S \ref{sec:SL3C} above, for the qutrit case, and following the standard description for qubits as reviewed in appendix \ref{sec:TwoQubitReview} below, local measurement protocols can include as special cases, more general types of symmetry group actions on quantum states and density operators than simply local unitary transformations. We examine here the role played by the local groups $SL(2,{\mathbb C})$ and $SL(3,{\mathbb C})$ respectively. 

The combinatorial representation and character theory used in the previous section on local unitary invariants, was indeed carried out by default already using the characters of these complex groups, exploiting the irreducibility of the fundamental representation. Thus the counts of 
independent LU invariants and of LSL invariants coincide in the cases involving pure states. However, the situation for these SLOCC LSL groups is different in their guise as `relativistic' groups for mixed qubit and qutrit systems. They have a presentation via homomorphic images, as real matrix groups of the appropriate dimension ($4\times 4$ and $9 \times 9$, respectively, so that they are subgroups of $GL(4,{\mathbb R})$ and $GL(9,{\mathbb R})$), acting linearly on the space of projective coordinates for the density operator. For qubits, this is simply the well-known 2:1 covering map, leading to the local isomorphism $SL(2,{\mathbb C})_{\mathbb R}\cong SO(3,1)$, and establishing $SL(2,{\mathbb C})_{\mathbb R}$ as the covering group of the Lorentz group. For qutrits, the situation is analogous and establishes a different local isomorphism: in this case, a 3:1 covering map between $SL(3,{\mathbb C})_{\mathbb R}$, and a nine dimensional matrix group, which we here denote $H_d(8,1)$. The dimension is written as $(8,1)$ to emphasize the noncompact nature of the group, by analogy with the nomenclature for noncompact orthogonal groups, and the subscript ${}_d$ relates to its definition as an invariance group for a particular tensor (the generalized $9\times 9\times 9$ $\widetilde{d}$ coefficient, as described in \S \ref{sec:SL3C} above). We have described such (generically) `non-classical' groups and their characters elsewhere \cite{FauserJarvisKing2006nbr}, and elaborated on some of the Schur function constructs, extending the classical techniques needed to manipulate formal characters, in \cite{fauser:jarvis:king:2010a:vertex} (see also \cite{fauser:jarvis:king:2013:ribbon}). We here reiterate the main results, in order to formulate counting rules for the relevant local invariants.

In the case of the orthogonal groups, the description of irreducible characters requires the introduction of Schur functions of orthogonal type \cite{littlewood1940}, denoted ${[}\lambda{]}$ for partition $\lambda$. A major result is the transcription between these symmetric functions and standard Schur functions $\{\lambda\}$, formalized as the so-called branching rule 
$
\{\lambda\} = \sum_{\delta \in {\mathcal D}} {[}\lambda/\delta{]}
$
expressing the fact that an irreducible character $\{\lambda\}$ of the general linear group reduces to a sum of irreducible orthogonal group characters,
resulting from skewing ${[}\lambda/\delta{]}$ for every $\delta$ belonging to particular infinite set ${\mathcal D}$ (all partitions with even row lengths). ${\mathcal D}$ has a more abstract characterization, that of a plethysm
${\mathcal D}= M_{(2)}\equiv \{2\} \underline{\otimes} M$ of the $S$-function $\{2\}$ (reflecting the symmetric metric tensor) by the formal infinite series of all one-part partitions (all symmetrized powers), $M = 1 + \{1\} + \{ 2\} + \{3\} + \cdots$.
The relation between standard and orthogonal type Schur functions can be written symbolically as
$\{\lambda\} ={[}\lambda/M_{(2)}{]}$.

For the case of formal characters of the matrix group $H_d(8,1)$, our work \cite{FauserJarvisKing2006nbr} leads to a parallel formulation. The character of the defining, nine-dimensional representation is denoted ${[\![}1{]\!]}$, and there is a class of characters ${[\![}\lambda{]\!]}$ corresponding to arbitrary partitions, with the difference that such characters are, in general, only indecomposable, not irreducible. The role of ${\mathcal D}=M_{(2)}$ is now played by the formal series
$M_{(3)} \equiv \{3\} \underline{\otimes} M$ (reflecting the totally symmetric invariant 3-fold tensor). The branching rule $\{\lambda\} ={[\![}\lambda/M_{(3)}{]\!]}$ again specifies indecomposable, not irreducible, characters in general. In the present case where $H_d(8,1)$ is locally isomorphic to $SL(3,{\mathbb C})_{\mathbb R}$, we do not expect such pathologies, but double- or triple-counting may still arise because of the presence of associated representations and modification rules (see below). The analysis of universal characters for groups of this type is incomplete and has only been pursued in a case-by-case manner \cite{FauserJarvisKing2006nbr}.
We can now formulate\\

\noindent
\textbf{Theorem 2: Counting LSL invariants for bipartite mixed systems}\\
The number $h_n$ of linearly independent mixed state LSL invariants at degree $n$ in the density matrix is 
\\[-.5cm]
\nopagebreak
\begin{description}
\item[(a) Qubits]
\[
h_n = \sum_{\sigma\vdash n, \ell(\sigma) \le 4} \left.{[} \sigma/M_{(2)}{]}\cdot {[} \sigma/M_{(2)}{]} \right.\bigg|_{{[}0{]}\cdot {[}0{]}}
\]
\item[(b) Qutrits (conjecture)]\footnote{
A proof of these formulae (for either qubits or qutrits) is as follows. Let ${\mathcal M}$ denote the appropriate symmetric plethysm series $M_{(2)}$ or $M_{(3)}$ and ${\mathcal L}$ its inverse. Then the inverse branching rule can be written ${[}\lambda{]} = {\{}\lambda/{\mathcal L}{\}}$, from which
$({[}\lambda{]}\cdot {[}\mu{]})\underline{\otimes}{\{}n{\}} = \sum_{\sigma,\tau}g^{{\{}n{\}}}{}_{\sigma,\tau}
{[}(\{\lambda/{\mathcal L}\underline{\otimes}\sigma)/{\mathcal M}{]}\cdot
{[}(\{\mu/{\mathcal L}\underline{\otimes}\tau)/{\mathcal M}{]}$. In the present cases $\{1/{\mathcal L}\} = \{1\}$,
$\{1\}\underline{\otimes}\{\sigma\}=\{\sigma\}$ and the inner product coefficient is $\delta_{\sigma,\tau}$ for partitions of weight $n$. }\\[-.5cm]
\[
h_n = \sum_{\sigma\vdash n, \ell(\sigma) \le 9} \left.{[\![} \sigma/M_{(3)}{]\!]}\cdot {[\![} \sigma/M_{(3)}{]\!]} \right.\bigg|_{{[\![}0{]\!]}\cdot {[\![}0{]\!]}}
\]
\mbox{}\hfill $\Box$
\end{description}
In contrast to the counting of local unitary invariants, here it is not possible to give a direct formula for the required multiplicity. The notation $\mbox{} |_{{[}\cdot {]}}$ indicates that the coefficient of the respective trivial character (the one dimensional representation), ${{[}0{]}\cdot {[}0{]}}$ or ${{[\![}0{]\!]}\cdot {[\![}0{]\!]}}$, should be extracted after the summation and skew operations have been carried out. The reason is that the specification of symmetric functions of orthogonal type 
${[}\lambda{]}$ contains redundant labelling, and a final stage
of \emph{modification} is required to determine if non-standard characters are either equivalent to standard ones, are zero, or are equivalent formally to the negative of standard ones. The final summation therefore contains possible alternating signs and internal cancellations, which need to be taken into account. The situation for characters of type ${[\![}\lambda{]\!]}$ is similar, but the result is formulated as a conjecture since the relevant modification rules are not known (see \cite{FauserJarvisKing2006nbr}).

\subsection*{$SL(2,{\mathbb C})\times SL(2,{\mathbb C})$ invariants for 2 qubit mixed systems}
Implementing the above theorem in the qubit case leads, as claimed in appendix \ref{sec:TwoQubitReview} below, to the Molien series
\begin{equation}
\label{eq:MolienLSLqubAll}
h(z) = 1+z^2 + 3z^4 + 4z^6 + 7z^8 + 9z^{10} + 14 z^{12} + \cdots = \frac{1+z^4}{(1-z^2)(1-z^4)(1-z^6)(1-z^8)}\,.
\end{equation}
With respect to the above discussion of character modifications, it is instructive to note how the $h_n$ coefficients are built. In the first place, only even partition weights arise because no $S$-function ${\{}\lambda{\}}$ of odd weight can skew with a plethysm of ${\{}2{\}}$ to give ${\{}0{\}}$. At weight 4 we have the obvious cases ${[}4/M_{(2)}{]}$,
${[}2^2/M_{(2)}{]}$ which skew to ${[}0{]}$ by the corresponding elements of 
$M_{(2)}$ given that ${\{}2{\} }\underline{\otimes}{\{}2{\}} =
{\{}4{\}}+{\{}2^2{\}}$. However, ${[}1^4/M_{(2)}{]}$ includes ${[}1^4{]}$ itself, and this 
non-standard character modifies to ${[}0{]}$ giving a total coefficient of 3, reflecting the existence of the alternative quartic forms $Q_4$ and $\widetilde{Q}{}_4$ (see appendix \ref{sec:TwoQubitReview} below).
At weight 6 we have similarly have the cases ${[}6/M_{(2)}{]}$,
${[}42/M_{(2)}{]}$ and ${[}2^3/M_{(2)}{]}$ which skew to ${[}0{]}$ by the corresponding elements of 
$M_{(2)}$; also ${[}31^3/M_{(2)}{]}$ contains ${[}1^4{]}$ after skewing by $\{2\}$,
which again modifies to ${[}0{]}$ giving a total coefficient of 4, rather than 3. Thus there is one additional algebraically independent invariant at degree 6. This pattern continues at degree $8$, 
but at degree 10, there is now a cancellation, saturating the ${{[}0{]}\cdot {[}0{]}}$ coefficient $h_{10}$ at 9 and preventing further independent invariants at this degree or higher (see also \cite{LuqueThibon2003pi4q}).

\subsection*{$SL(3,{\mathbb C})\times SL(3,{\mathbb C})$ invariants for 2 qutrit mixed systems}
As mentioned already, the character theory for groups such as $H_d(8,1)$ is not developed completely, and the count of invariants given above in this case should be taken subject to confirmation by explicit constructions. The lowest terms of the Molien series resulting from Theorem 2 are taken to be
\begin{equation}
\label{eq:MolienHd81app}
1+z^3+2 z^6+5 z^9+12 z^{12} +\cdots = \frac{1+ \cdots}{(1-z^3)(1-z^6)(1-z^9)^3(1-z^{12})^6\cdots}\,,
\end{equation}
simply reflecting the multiplicities at each weight occurring in the expansion of the relevant symmetric function series
\begin{align}
M_{(3)} =& \, \{0\} + \{ 3\} + \{3\}\underline{\otimes}\{2\} + \{3\}\underline{\otimes}\{3\} +\{3\}\underline{\otimes}\{4\} +\cdots
\nonumber \\
=& \, 
\{0\} + \{3\} + \big( \{6\} + \{42\}\big) +
 \big(\{9\} + \{72\} + \{63\} + \{52^2 \} + \{4^2 1\}\big) + \nonumber \\
& \, + \big(\{12 \} + \{10 \,2\} + \{93\} + \{84\} + \{82^2 \} + \{741\} + \{732\} + \nonumber \\
&\, + \{6^2 \} + \{642\}
       + \{62^3 \} + \{5421\} + \{4^3 \}\big) + \cdots\,,
\nonumber
\end{align}
on the plausible assumption that modification rules, and hence cancellations, will not affect these lowest degree counts.
As discussed in \S \ref{sec:SL3C} above, one invariant at each of degrees 3 and 6, three at degree 9, and 6 at degree 12 are expected. This count is partially confirmed in \S \ref{sec:SL3C} where the invariants at degrees 3 and 6 are given explicitly.
\section{Homogeneous polynomial entanglement monotones and mixed state systems.}
\label{sec:Monotones}
As mentioned in \S \ref{sec:SL3C} above, the identification of local unitary invariants (\S \ref{sec:LocalUnitary}) is only the first step towards useful entanglement measures. After quantum operations such as
(\ref{eq:RhoTransf}), such a quantity which is a function of the components of the density operator, say $f(\rho)$, is re-evaluated as $f(\rho')$. Since local measurements should not increase the degree of entanglement, the quantity $f$ must be an entanglement monotone, namely we must have the concavity condition
\[ 
p_1 f(\rho'_1)+ p_2 f(\rho'_2) \le f(\rho).
\]
Given a listing of unitary invariants such as Tables \ref{tab:QuadCub}, \ref{tab:GradedCount} above in the two qutrit case, it is in general a difficult problem to assemble properly behaving entanglement monotones (a discussion of the two qubit case is given in appendix \ref{sec:TwoQubitReview} below; the local unitary invariants are provided explicitly as Table 1 of \cite{KingWelshJarvis2007}). However, for quantities which are invariant under the appropriate `relativistic' transformation group, and which are homogeneous polynomials in the (projective) components of the density operator, a well-known construction exists \cite{dur2000three} (see also 
\cite{eltschka2012multipartite}). We now briefly describe how it can be adapted to the qutrit mixed state case.

Recall that the context of the monotonicity requirement is the quantum operation induced by local measurement operators, for example $E_1\otimes I$, $E_2\otimes I$ such that $E_1{}^\dagger E_1+  E_2{}^\dagger E_2=I$. Each $E_i$, $i=1,2$ is assumed to admit a singular value decomposition, such that $E_i = U_i D_i V_i$ and unitaries $U_i$, $V_i$ in fact with $V_1=V_2 \equiv V$ resulting from the constraint condition. Thus we have for some $0 \le a^2, b^2,c^2 \le1$
\[
D_1 = \left(\begin{array}{ccc} a&0&0\\ 0&b&0 \\ 0&0&c \end{array}\right),\qquad
D_2 = \left(\begin{array}{ccc} \sqrt{1-a^2} &0&0\\  0&\sqrt{1-b^2} &0\\  0&0&\sqrt{1-c^2} \end{array}\right)
\]
and in the nonsingular case $0 < a^2, b^2,c^2 < 1$we can write $D_i \equiv (d_i)^{\frac 13} \widehat{D}_i$ with $d_i =Det(D_i)$, namely
$d_1 = abc$, $d_2 = \sqrt{(1-a^2)(1-b^2)(1-c^2)}$ and $Det(\widehat{D}_i) =1$, that is, 
$\widehat{D}_i\in SL(3,{\mathbb C})$.

Consider the variation under this measurement operation of a homogeneous polynomial $f(\rho)$ which is invariant under local $SL(3,{\mathbb C})\times SL(3,{\mathbb C})$ transformations on the density operator, with degree of homogeneity $h$. We note\footnote{For ease of writing the tensor product is omitted, thus $E_i \rightarrow E_i\otimes I$, and so on.}
\begin{align}
f(\rho'_i) =f({E_i \rho E_i{}^\dagger}) = &\, f\Big(\frac{{d_i}^{\frac 23}}{p_i}U_i{\widehat{D}_i}V\rho V{}^\dagger \widehat{D}_iU_i^\dagger\Big)
= \left(\frac{{d_i}^{\frac 23}}{p_i}\right)^{\!\!h}f\big(U_i{\widehat{D}_i}V\rho V{}^\dagger \widehat{D}_iU_i^\dagger\big) \nonumber 
\end{align}
Now
\begin{align}
\qquad f\big(U_i{\widehat{D}_i}V\rho V{}^\dagger \widehat{D}_iU_i^\dagger\big)=&\,
f\big({\widehat{D}_i}V\rho V{}^\dagger \widehat{D}_i\big)=f\big(V\rho V{}^\dagger \big)=f\big(\rho\big)\, ,\nonumber
\end{align}
where local unitary invariance, local $SL(3,{\mathbb C})$ invariance, and again local unitary invariance, have been used in the respective simplifying steps. Thus we have
\[
p_1 f(\rho'_1)+ p_2 f(\rho'_2) = \left(p_1\left(\frac{{d_1}^{\frac 23}}{p_1}\right)^{\!\!h} + p_2\left(\frac{{d_2}^{\frac 23}}{p_2}\right)^{\!\!h}\right)f(\rho).
\]

This will be $\le f(\rho)$ if $f(\rho)\ge 0$, and also if the prefactor is $\le 1$. A guarantee of positivity is simply to adopt the absolute value $|f(\rho)|$ as the invariant; although nonpolynomial, the homogeneity is of course unaffected. The concavity of the final expression then depends on the evaluation of the inequality
\[
p_1\left(\frac{{d_1}^{\frac 23}}{p_1}\right)^{\!\!h} + p_2\left(\frac{{d_2}^{\frac 23}}{p_2}\right)^{\!\!h} \le 1
\]
which entails computing $p_1,p_2$ as weighted sums over the diagonal matrix elements of $D_1^2$, $D_2^2$ with unknowns $x,y, (1-x-y)$ arising from partial traces of products of the unitary operator $V$ with $\rho$ 
(compare \cite{dur2000three,eltschka2012multipartite}). 
The resulting inequality is a rational expression to be satisfied for all $0\le x,y,1-x-y\le 1$. It is not known at present for what range of values of $h$ this condition admits solutions. 

A weaker possibility, which still achieves an entanglement monotone, is to choose a power law scaling of $|f(\rho)|$ which avoids explicit evaluation of the $p_i$, namely $h=1$. This special value can of course be attained for any homogeneous polynomial 
invariant $f(\rho)$ of degree $h$, by taking the definitive entanglement measure to be $F(\rho) := |f(\rho)|^{\frac {1}{h}}$ from the outset. In this case, repeating the argument, we require for all $0< a^2,b^2,c^2 < 1$,
\[
a^{\frac 23}b^{\frac 23}c^{\frac 23}+ (1-a^2)^{\frac 13}(1-b^2)^{\frac 13}(1-c^2)^{\frac 13} \le 1\,.
\]
This condition can easily be verified with elementary algebra, after simplifying via the following composition of increasing functions to remove the fractional $\textstyle{\frac 13}$ exponents: exponentiation, taking the cube, and taking the logarithm. The singular case can be handled simply as the limit, where one or more of$a^2,b^2,c^2$ $\rightarrow 0,1$. In such cases, the above condition collapses to a trivial identity. In this way the monotonicity of 
$F(\rho)$ is established for all $0 \le a^2, b^2,c^2 \le1$.

This argument applies directly to the $SL(3,{\mathbb C})$ mixed qutrit invariants
identified in \S \ref{sec:SL3C} above: namely, we choose $|C_3|^{\frac 13}$ and $|C_6|^{\frac 16}$, respectively.
\section{Local unitary invariants for two qubit mixed states, and $SL(2,{\mathbb C})$ entanglement monotones.}
\label{sec:TwoQubitReview}
In order to provide a context for and contrast with our present analysis of the two qutrit system, we here present a brief review of the two qubit mixed system, based on our earlier analysis \cite{KingWelshJarvis2007}, and its extension to entanglement monotones which we report on here.

In the paper \cite{KingWelshJarvis2007}, a complete count, and explicit identification, of all algebraically independent local unitary $SU(2)\times SU(2)$ polynomial invariants was presented. This work confirmed previous computations, and also extended them in the sense that the complete structure of the ring of invariants was identified, together with extensive calculations to verify auxiliary polynomial relations (syzygies). The enumeration was checked both combinatorially using character methods (see appendix \ref{sec:Schurology} above), as well as directly via Molien's theorem. The Molien series
\begin{equation}
\label{eq:MolienSU2SU2}
h(z) = \frac{1+z^4 + z^5 + 3z^6 + 2z^7+ 2z^8 +3z^9 +z^{10} + z^{11} + z^{15} }
{(1-z)(1-z^2)^3(1-z^3)^2 (1-z^4)^3 (1-z^6)}
\end{equation}
indicates a rich variety of invariants, consisting of 10 polynomially independent quantities, with an additional set of secondary invariants, typically having a discrete spectrum (for the complete list see Table 1 of \cite{KingWelshJarvis2007}).  As emphasized in the main text, knowledge of local invariants is a necessary first step towards the identification of entanglement monotones. A complete algorithm for deriving all such quantities given a list of local unitary invariants is not known.

Nonetheless, appeal to a higher symmetry group action, namely the local SLOCC transformation groups 
in the guise of $SL(2,{\mathbb C})_{\mathbb R} \cong SO(3,1)$ acting as real linear tranformations on the space of projective coordinates of the density operator, allows a type of `Lorentz' singular value decomposition to be applied, and for canonical forms of quantities related to $\rho \widetilde{\rho}$ to be identified (see below). A complete diagonalization is not achievable under $SO(3,1) \times SO(3,1)$ local Lorentz transformations, with various exceptional classes in addition to the standard  forms \cite{avronbiskerkenneth2007v2qu,avronkenneth2009eg2qu}.

An alternative strategy, which we take up here, simply follows the line of identifying and constructing homogeneous polynomial invariants, this time of $SO(3,1) \times SO(3,1)$, and turns out to allow
certain types of entanglement monotones to be constructed. Although these quantities are not as fine-grained as the canonical forms, and may not distinguish exceptional types, they have the virtue of being easily computed and do not rely on carrying out a full diagonalisation. Moreover, as we indicate in \S \ref{sec:SL3C} above, they generalize easily to the qutrit case once the appropriate `relativistic' transformation group is identified.

The details are as follows. Recall that the single qubit density operator in the Pauli matrix presentation reads
\[
\varrho = \textstyle{\frac 12}{\mathbb I}_2 + \sum_{a=1}^3 r^a \sigma_a\,.
\]
The role of the Lorentz group arises from the observation that under invertible operations 
$\varrho \rightarrow \varrho' = A \varrho A^\dagger$, we have $Det(\varrho) = Det (\varrho')$ provided $A$ itself has unit determinant, that is $A \in SL(2,{\mathbb C})$. However, the determinant is easily seen to be
\[
Det(\varrho) = \textstyle{\frac 14} - \sum r^ar^a\,.
\]
Since $\varrho'$ no longer has unit trace, the transformation by $A$ must be accompanied by a re-scaling, and it is useful to append an additional projective coordinate ${\textsl r}^0$,
\[
\varrho = {\textsl r}^0{\mathbb I}_2 + \textstyle{\sum}_{a=1}^3 {\textsl r}^a \sigma_a \,.
\]
The determinant is now 
\[
Det(\varrho) = {\textsl r}^0{\textsl r}^0-\textstyle{\sum}{\textsl r}^a{\textsl r}^a \equiv
\textstyle{\sum}_{\alpha=0}^3 {\textsl r}^\alpha{\textsl r}^\beta \eta_{\alpha \beta}
\]
and so the group $SL(2,{\mathbb C})$ is seen as acting as a matrix group acting on the 4 dimensional space of 
projective coordinates of $\varrho$, and preserving the bilinear form ${\textsl r}^\alpha{\textsl r}^\beta \eta_{\alpha \beta}$ which is of course nothing but the standard Lorentz metric, with the matrix transformations identified with the Lorentz group $SO(3,1)$ in this case. The one- and two-qubit density operators are therefore
\begin{align}
\varrho = & \, \textsl{r}^0 {\mathbb I}_2 +  \textstyle{\sum}_{a=1}^3 {\textsl r}^a \sigma_a \equiv \textstyle{\sum}_{\alpha=0}^3
{\textsl r}^\alpha \sigma_\alpha\,, \nonumber \\
\rho = & \, \textsl{r}^{0\underline{0}} {\mathbb I}_4 +  \textstyle{\sum}_{a=1}^3 r^a \sigma_a\otimes {\mathbb I}_2 +
\textstyle{\sum}_{a=1}^3 r^{\underline{a}}{\mathbb I}_2 \otimes\sigma_a +
\textstyle{\sum}_{a,\underline{a}=1}^3 R^{a\underline{a}} \sigma_a\otimes \sigma_{\underline{a}} \,, \nonumber \\
 \equiv &\, \textstyle{\sum}_{\alpha, \underline{\alpha} = 0}^3 r^{\alpha\underline{\alpha}}\sigma_\alpha\otimes \sigma_{\underline{\alpha}} .
 \nonumber
 \end{align}
The above-mentioned Lorentz singular value decomposition proceeds with the analysis of the matrix 
$r (\underline{\eta}\hskip.1ex r^\top\hskip-.2ex \eta)$. This simply amounts to forming the tensor 
$w^\alpha{}_\beta=r^{\alpha\underline{\alpha}}\eta_{\underline{\alpha}\underline{\beta}}r^{\gamma\underline{\beta}}
\eta_{\gamma\beta}$, which evidently transforms only under one local Lorentz group, by the usual rules of raising, lowering and contraction of indices (there is an equivalent tensor $w^{\underline{\alpha}}{}_{\underline{\beta}}$ which is isospectral, and transforms under the other local Lorentz group). As mentioned, a complete diagonalization is  achievable in all but some exceptional cases, and the analysis is consistent with the emergence of the well-known convex roof extension entanglement measure \cite{wootters1998entanglement}
$max(\lambda_1^\downarrow-\lambda_2^\downarrow-\lambda_3^\downarrow-\lambda_4^\downarrow,0)$.

From the perspective of group representations, as emphasized in this paper, the problem is again the enumeration and construction of polynomial invariants of the local group $SO(3,1)\times SO(3,1)$, for the representation corresponding to $r^{\alpha\underline{\alpha}}$, namely the direct product of defining 4-dimensional representations (with character ${[}1{]}\cdot {[}1{]}$ in standard notation; see appendix \ref{sec:Schurology} above). For finite dimensional representations this can be carried out for the corresponding problem in $SO(4)\times SO(4)$, and via the isomorphism $SO(4)\cong SU(2)\times SU(2)$ the count becomes combinatorially identical to the problem of identifying local unitary invariants for the \emph{four} qubit \emph{pure state} system \cite{LuqueThibon2003pi4q} (see also \cite{levay2006geometry}). The Molien series can be readily computed directly from Molien's theorem in this case, or evaluated using character manipulations (see Theorem 2, appendix \ref{sec:Schurology} above); 
counting invariants up to degree 12 gives 
\begin{align}
h(z) = & \,1+z^2 + 3z^4 + 4z^6 + 7z^8 + 9z^{10} + 14 z^{12} + \cdots 
\label{eq:MolienLSLqubTerms} \\
& \, = \frac{1}{(1-z^2)(1-z^4)^2(1-z^6)}\equiv  \frac{1+z^4}{(1-z^2)(1-z^4)(1-z^6)(1-z^8)}
\label{eq:MolienLSLqubRatl}
\end{align}
which is consistent with an invariant ring generated by the independent invariant traces $Q_{2p}=Tr(w^p)$ of the matrix $w$, $p=1,2,3,4$, with an additional constraint between $Q_8$ and the square of the determinant, $\widetilde{Q}_4 = Det(\rho)$, which plays the role of a secondary invariant in this case (see also \cite{KingWelshJarvis2007}).

As an illustration of the method, we evaluate the lowest degree invariants,
$Q_2$, $Q_4$ and $\widetilde{Q}_4$, in order to show explicitly how combinations of local unitary invariants combine in forming these quantities. Using the transcription between the basis $\{ r^a, r^{\underline{a}},
R^{a\underline{a}} \}$ and the Lorentz-covariant set $\{ r^{\alpha \underline{\alpha}}\}$ used above, and setting 
$\{r^{0\underline{0}}\}$ to its standard value $r^{0\underline{0}} = \textstyle{\frac 14}$, we compute
\begin{align}
Q_2=Tr\big(w\big) = & \, r^{\alpha\underline{\alpha}}\eta_{\underline{\alpha}\underline{\beta}}r^{\gamma\underline{\beta}}
\eta_{\gamma\alpha} =
(r^{0\underline{0}})^2 - (r^{a \underline{0}})^2 -
(r^{0 \underline{\alpha}})^2 + (r^{a \underline{\alpha}})^2 
\nonumber \\
\equiv & \,  \textstyle{\frac{1}{16}} - r^2 -\overline{r}^2 +  RR,
\end{align}
where the final line gives the expansion in terms of the list given in Table 1 of \cite{KingWelshJarvis2007} using the 
obvious notation $r^2 =  \sum_a r^a r^a$, $\overline{r}^2 = \sum_{\overline{a}} {r}^{\overline{a}}{r}^{\overline{a}}$,
$RR = \sum_{a,\overline{a}}R^{a\overline{a}} R^{a\overline{a}}$.
As can be seen, the resulting form is \emph{inhomogeneous} in the underlying local unitary invariants because of the requirement of trace normalization. This situation continues with the remaining forms, the first of which is
\begin{align}
Q_4=Tr\big(w^2\big) = & \,\sum r^{\alpha \underline{\beta}}r_{\beta \underline{\beta}}r^{\beta\underline{\gamma}}
r_{\alpha\underline{\gamma}} \nonumber \\
= & \, 
RRRR +(r^2)^2 + (\overline{r}^2)^2 - 2 r RRr -2 \overline{r}RR\overline{r} + rR\overline{r} 
-\textstyle{\frac{1}{8}} r^2 -\textstyle{\frac{1}{8}} \overline{r}^2 + \textstyle{\frac{1}{256}}\nonumber \\
\equiv & \,
K_7 + (K_3)^2 + (K_4)^2 -2(K_8+K_9) + K_6-\textstyle{\frac{1}{8}}(K_3+K_4)
+ \textstyle{\frac{1}{256}}, 
\end{align}
using the additional abbreviations $r RRr = \sum_{a,b,\overline{a}} r^a R^{a\overline{a}}R^{b \overline{a}}r^b$,
$\overline{r}RR\overline{r} = \sum_{a,\overline{a}, \overline{b}} r^{\overline{a}} R^{a\overline{a}}R^{a \overline{b}}r^{\overline{b}}$, $RRRR = \sum_{a,b,\overline{a},\overline{b}}R^{a\overline{a}}R^{b\overline{a}}
R^{b\overline{b}}R^{a\overline{b}}$. Introducing also 
$r (R \times\hskip-2.6ex\times R) \overline{r} =
\varepsilon^{ijk}  \varepsilon^{\underline{p}\underline{q}\underline{r}} r^{i} 
R^{j\underline{p}} R^{k\underline{q}}r^{\underline{r}}$ we have further
\begin{align}
\widetilde{Q}_4 := Det(\rho) = & \, 
\textstyle{\frac{1}{24}}\varepsilon_{\lambda \mu\rho\sigma} 
\varepsilon_{\underline{\alpha}\underline{\beta}\underline{\gamma}\underline{\delta}}
r^{\lambda\underline{\alpha}}r^{\mu\underline{\beta}}r^{\rho\underline{\gamma}}r^{\sigma\underline{\delta}}\nonumber \\
= & \, \textstyle{\frac 16}\varepsilon^{\underline{0}\underline{p}\underline{q}\underline{r}}
\Big(\varepsilon_{0ijk} r^0{}_{\underline{0}} r^i{}_{\underline{p}} r^j{}_{\underline{q}} r^k{}_{\underline{r}} +
\varepsilon_{i0jk} r^i{}_{\underline{0}} r^0{}_{\underline{p}} r^j{}_{\underline{q}} r^k{}_{\underline{r}}+ 
\varepsilon_{ij0k} r^i{}_{\underline{0}} r^j{}_{\underline{p}} r^0{}_{\underline{q}} r^k{}_{\underline{r}}+
\varepsilon_{ijk0} r^i{}_{\underline{0}} r^j{}_{\underline{p}} r^k{}_{\underline{q}} r^0{}_{\underline{r}} \Big)\nonumber\\
= & \,  
\textstyle{\frac 16}\Big( 
r^{0\underline{0}}\varepsilon^{ijk} \varepsilon^{\underline{p}\underline{q}\underline{r}}
r^{i\underline{p}} r^{j\underline{q}} r^{k\underline{r}} 
+3\varepsilon^{ijk}  \varepsilon^{\underline{r}\underline{p}\underline{q}} r^{i\underline{0}} 
r^{0\underline{r}} r^{j\underline{p}} r^{k\underline{q}} \Big)\nonumber \\
=& \, \textstyle{\frac {1}{4}}Det(R) + \textstyle{\frac 12}r (R \times\hskip-2.6ex\times R) \overline{r}
\nonumber \\
\equiv & \, \textstyle{\frac {1}{24}}K_5 + \textstyle{\frac 12}U_1\, .
\end{align}
We do not give the corresponding expression for $Q_6$ here, but it is clear that it can be expanded in an analogous manner.
Following the derivation in appendix \ref{sec:Monotones} above, these $SL(2,{\mathbb C})\times SL(2,{\mathbb C})$ invariants will provide entanglement monotones by taking appropriate powers of their absolute value, namely $|Q_2|^{\frac 12}$,  $|Q_4|^{\frac 14}$, $|\widetilde{Q}_4|^{\frac 14}$ and similarly $|Q_6|^{\frac 16}$.

\end{appendix}
\newpage

{\small
%
	{\small

%
%
\vfill
\tableofcontents
}
\end{document}